\documentclass{article}

\PassOptionsToPackage{numbers,compress}{natbib}
\usepackage[preprint]{neurips_2020}

\usepackage[utf8]{inputenc} 
\usepackage[T1]{fontenc}    
\usepackage{hyperref}       
\usepackage{url}            
\usepackage{booktabs}       
\usepackage{amsfonts}       
\usepackage{nicefrac}       
\usepackage{microtype}      
\usepackage{amsthm}         

\usepackage{graphicx}

\usepackage{caption}
\usepackage{tabularx}
\usepackage[skip=1ex]{caption}

\newcolumntype{C}{>{\centering\arraybackslash}X}

\newcommand\cmt[1]{}  
\newcommand\mymark[1]{}  

\theoremstyle{definition}
\newtheorem{definition}{Definition}[section]
\theoremstyle{observation}
\newtheorem{observation}{Observation}[section]

\title{Applied Awareness: Test-Driven GUI Development using Computer Vision and Cryptography}

\author{%
    Donald Beaver\thanks{%
        \texttt{don.beaver@gmail.com}, \url{https://sites.google.com/view/donbeaver}} \\
        Independent Scholar\\
        Pittsburgh, PA, USA
}

\begin{document}

\maketitle

\begin{abstract}

    Graphical user interface testing is significantly challenging, and automating it even more
    so.  Test-driven development is impractical: it generally requires an initial
    implementation of the GUI to generate golden images or to construct interactive test
    scenarios, and subsequent maintenance is costly.  While computer vision has been applied to
    several aspects of GUI testing, we demonstrate a novel and immediately applicable
    approach of interpreting GUI presentation in terms of backend communications, modeling
    "awareness" in the fashion employed by cryptographic proofs of security. This focus on
    backend communication circumvents deficiencies in typical testing methodologies that rely
    on platform-dependent UI affordances or accessibility features. Our interdisciplinary work
    is ready for off-the-shelf practice: we report self-contained, practical implementation
    with both online and offline validation, using simple designer specifications at the outset
    and specifically avoiding any requirements for a bootstrap implementation or golden images.
    In addition to practical implementation, ties to formal verification methods in
    cryptography are explored and explained, providing fertile perspectives on assurance in UI
    and interpretability in AI.

\end{abstract}


\section{Introduction}

User interfaces are often the hardest part of a system to test and maintain, particularly in
the face of rapid architectural changes, design updates, and disparate, evolving, distributed
backend services.  This work describes techniques for validating graphical user interfaces by
applying image classification to automate the interpretation of UI snapshots in terms of
underlying data model transmissions. Specifically: to be validated, a user interface rendering
must be aware of the underlying communications in a particular, technical sense.

This work explores and exhibits several aspects of validating GUIs using ML and crypto:
\begin{itemize}
\item Automated testing
\item Test-driven development without requiring baseline implementation
\item Simple, practical, performant implementation
\item Online and offline validation
\item Platform independence
\item Formal notions arising in cryptographic protocol validation
\end{itemize}

We describe an end-to-end implementation of a client-server based GUI validated by training CV
object detection to identify essential affordances and then create a convincingly fake version
of the JSON tree served by the backend.  When the fake JSON diverges from the actual JSON, a
validator reports GUI failures.  The trained ML system is small and fast enough for a GUI to be
"self-aware" of its performance: the GUI (optionally) shows its shame immediately, when it
realizes the JSON from its images doesn't match the JSON from the backend.

Remarkably, training on designer mockup slides in Keynote/PowerPoint suffices, prior to writing
any code. The validation paradigm is not only sufficient for automated testing, it makes
properly-disciplined test-driven development possible from scratch - where the UI tests are
written in advance of the first line of code. No ``golden images'' are needed; no fine-grained
anticipation of the deployment platform (iOS/Android iconography, map tile sources, etc.) are
needed.

\subsection{Paradigm summary}

There are many automated approaches to UI testing, including applying computer vision
techniques to determine quality of a result or to identify affordances in order to create
interaction sequences that better cover the application state space
\citep{ycm09-sikuli,cym10-gui-cv,nrbm14-guitar,qn13-survey,reg07-petri,bmars20-petri}.
These generally seek to compare the final rendering to an expected rendering - after code
has been changed, or after states have been traversed.  Namely, they focus on the presented
images and affordances.

In contrast, this work focuses on the communications model behind the scenes.
Taking a page from cryptographic protocol verification, we advance the following paradigm:

\begin{quote}
    A user interface enjoys {\em awareness-based assurance} if it is technically aware
    of its backend communications:
    an independent, automated interpreter should map the rendering to the
    backend communication stream faithfully enough to convince an automated validator.
\end{quote}

We give a concrete demonstration here, but this work is not so much about ``the most optimized
or robust object detector'' or ``the best test set'' as it is about the importance, relevance
and ease of using computer vision to give automated test assurance by way of ``awareness.''

\textbf{Roadmap:} \S\ref{s-background} Background;
\S\ref{s-defs} Awareness;
\S\ref{s-case-study} Case Study;
\S\ref{s-crypto} Crypto.


\section{Background} \label{s-background}

\subsection{Automated testing}

Quality assurance performed by humans can be tedious, error-prone and expensive. While
obtaining a baseline validation of an initial software deployment may be straightforward, the
effort in repeating QA after minor or major changes can be costly and tedious.  Automated test
suites are widely relied on for non-UI production software, but automated UI testing remains
brittle for many reasons.

First, even where automated UI testing is practiced, a very common approach is the "golden
image" approach: compare images after a change to already-validated images stored prior to a
change.  Acceptance criteria may demand the comparison be pixel perfect, or it may allow for
deviations measured by image distance or similar treatments. Often, a human needs to be
bothered to eyeball differences. When discrepancies occur, new snapshots are often trusted if
they ``look good.''  While images are small in modern storage terms, they are unwieldy in
source control repositories tailored to track text changes.

Second, some UI testing approaches are particularly platform-dependent, requiring extensive
plaform expertise.  Apple's iOS/MacOS platform is exceptional in its robust accessibility
features - and these features have a dual use to support UI testing. The UI components are
already instrumented and tailorable for alternate interactions: for example, using voice to
trigger a button action rather than requiring a screen tap; or having explicit metadata
associated for vocalizing an otherwise-visual presentation. Apple's goodwill provides a
significant development benefit: one can write UI tests that take advantage of accessibility
features to track and assert expected affordances. This approach has such immediate benefit
that Apple's XCode now offers up templates and guides \citep{xcode}.  But it requires a
specific platform and a reasonably experienced developer.

Third, somewhat more principled and abstract (and debated), there is the pitfall of testing the
implementation rather than the contract. In ordinary testing, one should not be examining
whether a square root algorithm is faithfully following Newton's algorithm. One should just
verify if the result, squared, is expected. For us, the presence of a specific DOM element in a
renderer model is not necessarily the goal: the ability to resurrect the conveyed backend
messages from such elements is the "contract" to be vetted.  Certainly, one can be interested
in both - but we are not focusing on DOM elements.  DOM elements or UIKit views are
characteristics of the current "internal" implementation of a rendering, not of the "affordance
contract."

Finally, navigating the state transitions of a UI is a distinct challenge, particularly for
generating broad case coverage.  There is an established line of research into automatically
identifying affordances and generating user interactions to trigger comprehensive state changes
\citep{reg07-petri, ycm09-sikuli,cym10-gui-cv, gcs13-grammat,qn13-survey, nrbm14-guitar,
bmars20-petri}.  In fact, a state-change bug was the original motivation for the current work,
although our focus is not specifically on finding or triggering state changes.

State transitions are important but also somewhat complementary in our setting. We require
validations be satisfactory after sequences of state transitions, but we are focused on how the
validation occurs once those state transitions have been triggered. We seek to expand the menu
of complementary options available to applying CV to testing.  With our ability to enable
online validation, we can also leverage live feedback in A/B testing, relying on empirical
behaviors to validate the important regions of state space.

In sum, there are many concerns or gaps in preceding work, which we propose to circumvent or
fill: \begin{itemize} \item Brittleness \item Difficult maintenance \item Platform-dependent
        tricks \item Delayed, offline image/log aggregation and pipelining \item Subjective
        aspects of automated difference-checking \item Narrow focus on the renderings
\end{itemize}

\subsubsection{Test-driven GUI development}

Automated TDD for GUIs -- as distinguished from "testing GUIs after the fact" -- is extremely
challenging.  Tests for DOM elements could be written up front, but this flavor is "testing the
implementation" rather than "testing the API."  Tests for images invariably require a bootstrap
implementation.  Even with a bootstrapped set of images, moving forward in a TDD manner is
quite hard, if ever actually performed in a principled TDD way. After moving a button, for
example, generating the next round of valid images (or image validators) is vastly easier using
approved logs from already-modified code than editing previous golden snapshots.

Turning our attention away from image fidelity or DOM fidelity, and focusing it on the
communications being represented by the GUI, we can both theoretically and practically produce
tests and validation methods up front, before any code is written or changed.  We don't need to
run square-root algorithm to build our square-root test cases; we don't need to count the
expected Newton iterations to have confidence.

ML-based image analysis is the key to obtaining sufficiently general flexibility. It is,
arguably, what the human does, when choosing to accept a new golden snapshot. The human
verifies that the new snapshot is sensible, in that it expresses states conveyed by backend
messaging.

Naturally, object detection and the human can also verify specific design aspects, such as "the
button is 44x44 pixels" - but this is already relatively covered ground.  We are focused on the
"semantics" of the image as explained by the backend communications.  The JSON offers a
testable interpretation of what the GUI is "aware" of representing.

\subsection{Cryptography and Simulated Environments}

Relationships to validation of cryptographic protocols are described in detail in
\S\ref{s-crypto}.  Briefly, prominent formal methods require of course that an attacker finds a
cryptosystem hard to decipher, but they demand more strictly that an adversary can only send
encryptions using a cryptosystem if the adversary is {\em aware} of the cleartext content. This
subtle demand for awareness is decades old, but it also took decades to emerge, and it is
critical in formal security analysis.

The rendering of UI is analogous to encryption of a cleartext. The UI is intended for "easy
deciphering" by humans, of course. Here, as in cryptosystems, we demand that the UI is
mechanically decipherable to reveal the messages employed in its formulation. When this demand
for awareness is satisfiable -- and we show it is, in \S\ref{s-case-study} -- the "deciphering"
proves useful as a tool for validating the UI renderings ({\em not} in terms of other
renderings but in terms of cleartexts).

Furthermore, drawing on cryptographic principles also provides critical notions and pitfalls
when analyzing aspects individually.  Concretely, consider showing that a GUI's renderings are
aware of the left branch of JSON messages, and then showing that the renderings are aware of
the right branch of JSON messages.  One would like to conclude that the composition is
therefore validated.  In cryptography, however, the combination of properties like integrity
and confidentiality are sufficiently subtle that naive composition can lead to failures in
validation if not outright breakage.  Any approach for composing GUI validation efforts does
well to be informed by the potential for subtle gaps.


\section{Synthesizing Backend Communications from GUI Renderings} \label{s-defs}

The formalism here is more to pin down what we mean and to tie it in with cryptographic
validation paradigms later. For the most part, we are soon going to focus on achieving it in
practice.

Let's restrict our attention to a \textit{standard architecture} containing a frontend with a
client receiving a data model comprised of a text-based dictionary tree, received from a
backend service on a regular basis.  The frontend GUI $G$ renders images $G(x)$ based on the
supplied tree $x$.  The following formalization captures the notion of \textit{awareness of the
input} by demanding the ability to recover the input from the rendering:
\begin{definition} \label{def-aware}
    A GUI $G$ is $\epsilon$-\textit{aware} of the model, with respect to an input distribution
    $D$ on model data, if there exists an efficient interpreter $I$ such that $I(G(x)) = x$
    with probability exceeding $1-\epsilon$.
\end{definition}
Typically, there will be information unrepresented in the GUI, so let's allow for a filter $F$
to confine our demands to a subset of the model data:
\begin{definition} \label{def-aware-filter}
    A GUI $G$ is $\epsilon$-\textit{aware} of the model \textit{relative to filter} $F$
    if $F(I(G(x))) = F(x)$ under conditions from \ref{def-aware}.
\end{definition}

The preferred distribution is what is seen in the deployment in the wild, of course, but this
is hard to track. We break the domain into discovering desirable distributions versus measuring
awareness for a given distribution, and focus on the latter.  The job of discovering good test
coverage (for implementation logic and for domain examples) is critical, of course, but our
focus will be on what "awareness" means and why it is helpful when we have a distribution
already in mind.

\begin{definition}
    Let a test suite $S$ generate backend text models with distribution $D$.
    A GUI $G$ satisfies \textit{awareness-based assurance} with respect to test suite $S$
    and filter $F$ if it is $\epsilon$-aware of the model with respect to $D$ and
    relative to $F$.
\end{definition}

In more concrete and computational terms, let's phrase this in terms of distinguishing power.
We would like to have a validator $V$ who, given filtered views $F(I(G(x))$ and $F(x)$, cannot
determine which is the actual input and which is the interpretation of its rendering.

\begin{definition}
    Let $V$ take an input pair of strings, $(a,b)$, and report a 1-bit output.
    A test validator $V$ \textit{demonstrates awareness-based assurance} of GUI $G$
    if $\left| Pr[V(f_0, f_1) = i] - 1/2 \right| < \epsilon$, where $i$ is a uniformly
    random bit, $f_i$ is $F(x)$ with $x$ sampled from $D$, and $f_{1-i}=F(I(G((x)))$.
\end{definition}

The validators are largely very simple, generally just being an equality test of a filtered
branch of a tree.

\subsection{Timeseries}


A few brief remarks on an important generalization, without diving too much farther into
formalization for its own sake. In practice, we encounter a log of GUI snapshot images and a
log of backend text messages.
Ideally these have near identical timestamps and can be easily
correlated, so that the pair

of model and rendering-interpretation can be supplied to the
Validator.  A GUI rendering might lag the backend message badly, such that another backend
message arrives before the rendering finishes. Logs can be recorded by distinct processes and
at distinct rates, particularly when log image storage is costly. For a given logged image, we
will simply pair it with the most recent logged backend message preceding it.  Robustness and
error-tolerance of the overall validation could be improved by allowing leeway in this pairing.
For example, we can lower the bar so that the rendering-interpretation matches any one of a
window of backend messages. Our empirical explorations don't need this, but some may.


\section{Case Study} \label{s-case-study}

We demonstrate the paradigm by implementing a GUI from scratch, starting with "designer spec,"
and coding the GUI and backend independently of the bootstrap object detector and validator.
Even very recently these efforts suffered from cross-platform variation: training models and
doing validation on Linux, while running the GUI and aggregating screenshots in iOS, for
example. Python, TensorFlow or PyTorch, ObjectiveC or Swift: many parts to master.
A single platform is not {\em required} but it enables substantially easier
implementation -- and even online validation: the GUI can test itself live.

In our setting, we used Swift on a MacBook Pro, end-to-end, to implement GUI, backend model
messages, logging, testing (offline, and live online self-testing), and object detection
training and evaluation. For initial design spec: Keynote; and Preview for snapshots and
labels.

\subsection{Test-Driven Design Phase}

Our sample application is a GUI for a drone flying over a mapped area. We wish to be assured
that warning conditions are exhibited correctly, e.g. when flying in risky or dangerous
circumstances. This is a simplified example motivated by an actual production setting (and on
expensive debugging of a flaw), but the production code wasn't itself TDD-developed nor can it
be released. The spec and code here will be available on github.

Disciplined TDD requires test cases prior to writing code. We approach this via straightforward
steps:

\begin{enumerate}
    \item Designer provides a slide deck spec (e.g. Keynote; any image-generating tool is fine)
    \item Backend model representation is defined (JSON schema)
    \item An object detector is trained on labeled spec (design deck) to identify affordances
    \item Test code maps presence (or absence) of affordances to synthesized JSON
\end{enumerate}

The idea of identifying affordances via CV is not remotely novel: many predecessors such as
Sikuli \citep{ycm09-sikuli} and others have used it for years. Those prior applications focused
on the challenging task of automatically generating test sequences and comparing images. Here,
instead, we focus on awareness of the backend messages.

\begin{table}[t]
    \centering
\begin{tabularx}{\columnwidth}{CC}
    \includegraphics[scale=0.13]{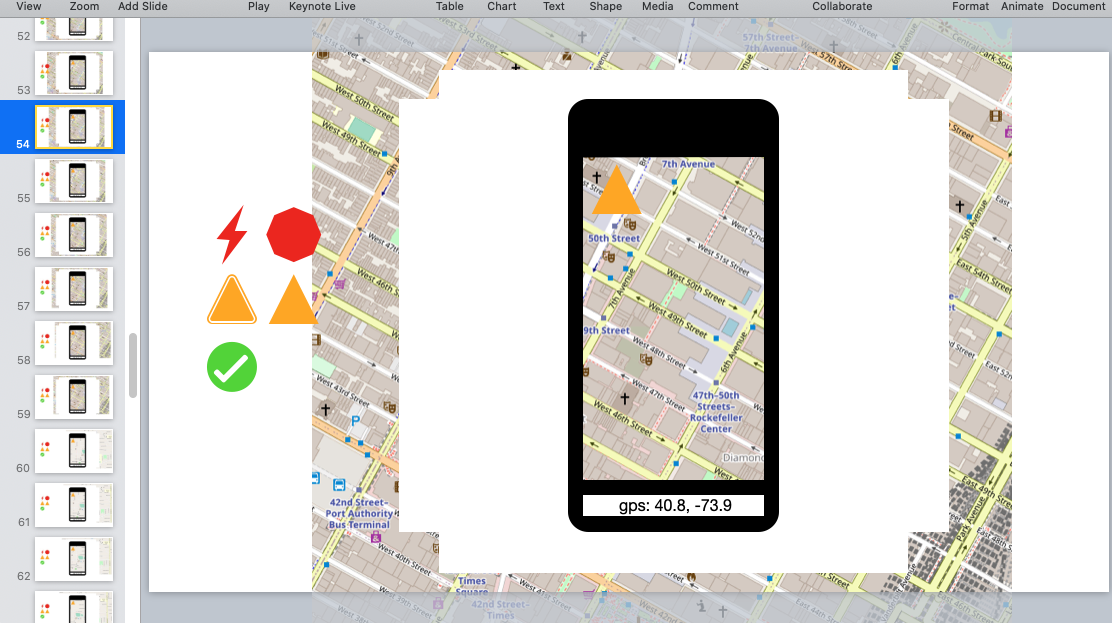}
    \captionof{figure}{Designer ``tool'' - Apple Keynote slide,
    simple icon palette, draggable Open Street Maps background image.}
    \label{fig-design-tool}
    &   
    \includegraphics[scale=0.13]{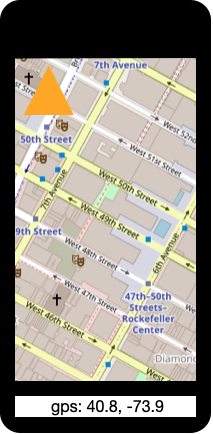}   
    \captionof{figure}{Screenshot of design (specifically not a golden image),
    for training the validator.}
    \label{fig-design-train}
\end{tabularx}
\end{table}%

The design spec is a Keynote deck starting with 3 slides as visual spec but expanded to 90
slides as synthetic test cases; 30 for each warning condition:
    {\em nominal, caution, danger}.

The Keynote design spec used a Keynote-provided standard iPhone wireframe and Open Street Maps
images for background; see fig.~\ref{fig-design-tool}.

Synthetic images were generated and labeled manually in Keynote (fig.~\ref{fig-design-train}),
by dragging background maps and updating affordances.  Although tedious, snapshotting took 10
seconds per image, 45 minutes for the full training/test set to support awareness-based TDD.
The background variations were sufficiently diverse for robust training.  Fully synthetic
generation is completely compatible with this approach, of course, whether in ``test-driven'' or
``post-testing'' scenarios.

A CoreML (Apple) object detection model was trained on 250 iterations using transfer learning
taking 78 minutes on a 2016 MacBook Pro.  60 images were used for training and 30 held for test.
The resulting model was 61MB in size.

Results of later evaluation can be seen in fig.~\ref{fig-eval-caution}, where we see that the
model trained on the designer spec is capable of locating the affordance and classifying it
correctly.

\begin{table}[t]
    \centering
    \includegraphics[scale=0.13]{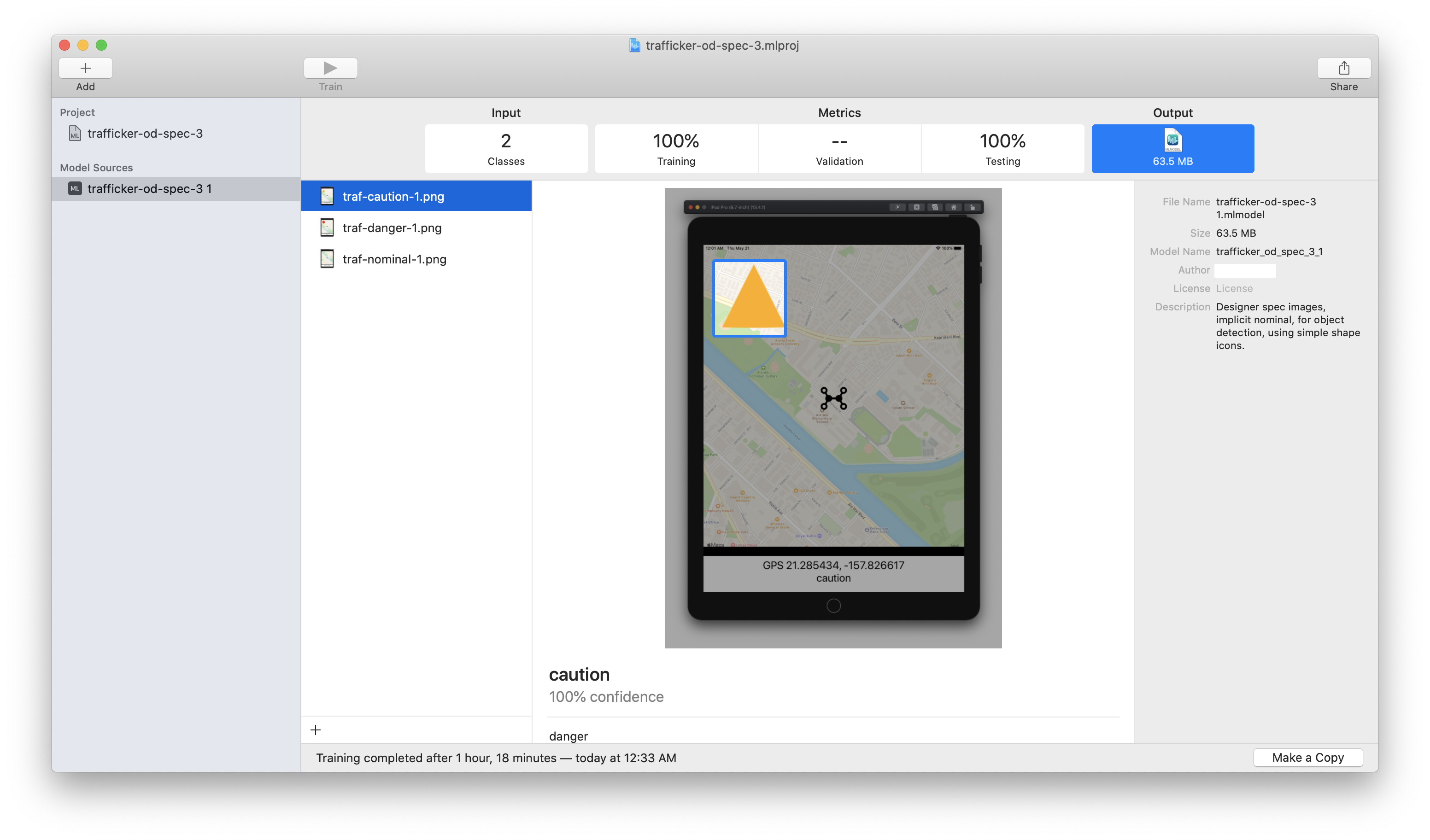}
    
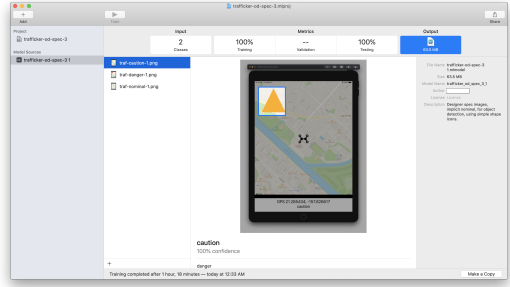
\captionof{figure}{Evaluation of GUI snapshot using OD model trained on designer spec.}
    \label{fig-eval-caution}
\end{table}%

\subsection{Backend and Frontend Coding}

The backend and frontend code are written and subjected to the testing above.  In strict TDD,
the tests are constructed first: here, rough designer images sufficed.  Of course, for
post-test scenarios, images from an existing production implementation or MVP are
straightforward to label automatically using the backend JSON, to train for forward awareness
after revisions.

The GUI was presented as a 9.7-inch iPad whose aspect ratio and dimensions differ from the
design spec (iPhone). The GUI rendered Apple Map tiles, easily distinguishable from designer
spec OSM images.

A smoke test was immediately successful: the first renderings of the GUI were evaluated using
the object detection model to verify detections.  Identification was comprehensive -- using OD
made variations in size, map tiles, and other affordances irrelevant for testing.

The Validator logic consisted of filtering the warning mode from the JSON tree and making sure
that the rendering-interpretation matches the provided JSON. Specifically and trivially:
evaluate the object detector, build JSON, all within XCode \citep{xcode}.

The code for logging GUI images, for logging JSON messages, for pairing up logged GUI images
with most recent (independently-logged) JSON backend messages, is straightforward and occupied
as much of the validation effort from scratch as the JSON-interpretation itself.  Naturally,
more complex settings would require greater effort -- for example, supporting different
platforms, or reverse-estimating GPS from maps.  This app instrumentation is a fixed overhead,
however; many different facets of a GUI and JSON tree can be explored using the same
infrastructure.

\begin{observation}

    The backend, GUI, awareness-interpreter and validator (all to be released on github)
    demonstrate a feasible implementation of test-driven GUI development using automated
    awareness-based validation, relative to a filter of JSON on the warning mode and with
    respect to a test suite comprised of pseudorandomly generated GPS waypoints over an urban
    area.

\end{observation}

\subsection{Debugging and Fault Injection} \label{subsec-debug}

%

\begin{table}[ht]
    \centering
\begin{tabularx}{\columnwidth}{CC}
    \includegraphics[scale=0.13]{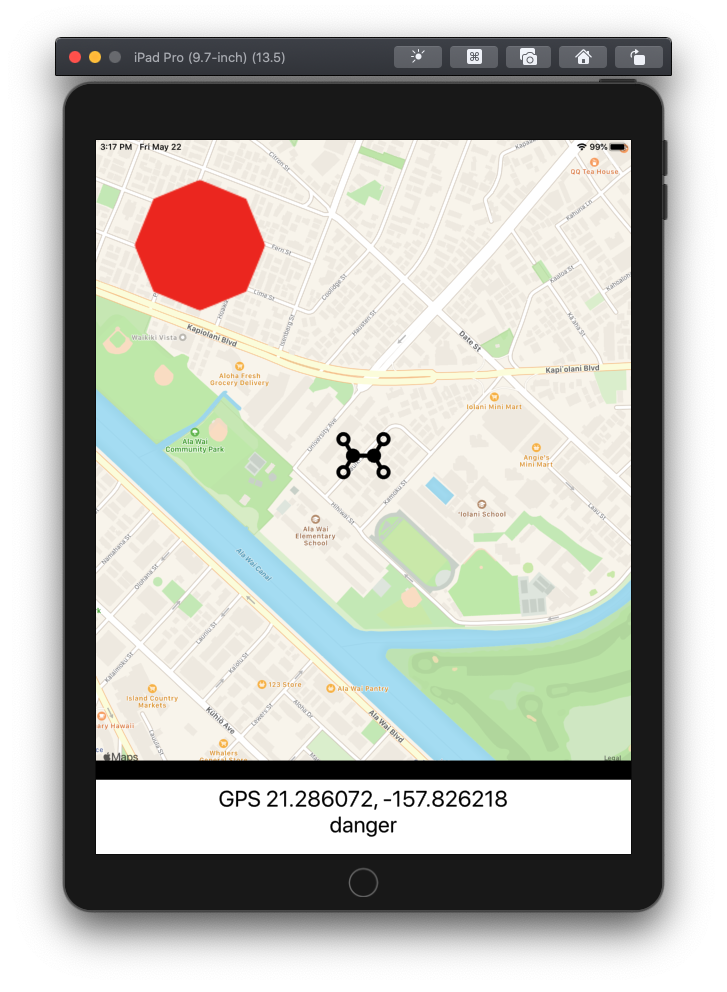}
    \captionof{figure}{Correctly-implemented GUI shows ``danger'' affordance
    based on backend JSON.}
    \label{fig-correct-gui}
    &   
    \includegraphics[scale=0.13]{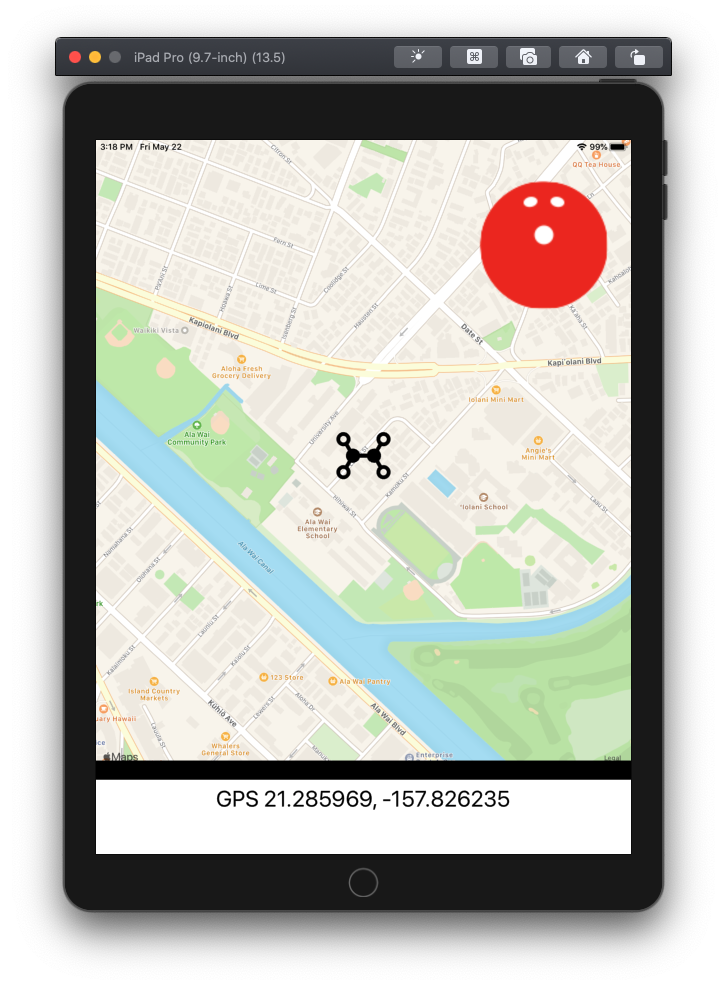}   
    \captionof{figure}{After fault-injection, self-aware GUI shows shame when
    \texttt{warningMode} JSON interpreted from current screenshot
    (lacking required danger-affordance) fails to match actual backend JSON.}
    \label{fig-inject-shame}
    \\
    \begin{tabular}{ll}
        {\small perceived JSON:} & \texttt{\small \{"warningMode":2, ...} \\
        {\small backend JSON:}    & \texttt{\small \{"warningMode":2, ...} \\
        ~
    \end{tabular}
    \captionof{figure}{Interpretation of rendering demonstrates awareness of backend JSON.}
    \label{fig-aware-good}
    &   
    \begin{tabular}{ll}
        {\small perceived JSON:} & \texttt{\small \{"warningMode":0, ...} \\
        {\small backend JSON:}    & \texttt{\small \{"warningMode":2, ...} \\
        ~
    \end{tabular}
    \captionof{figure}{Interpretation of faulty rendering correctly demonstrates
    a gap in awareness.}
    \label{fig-aware-bad}

\end{tabularx}
\end{table}%

We validated the validation by injecting faults.
We injected different kinds of failures but will present a particular one here.
Common UI implementations first check for changes in latest JSON and trigger re-rendering only
when change in the whole tree or a subscribed branch is detected.
Motivated by a real-life example, we injected a failure to note transitions between
{\em caution} and {\em danger}.  One can imagine that changing a JSON model from one version to
the next leads readily to these failures.  A ``nominal'' state might be represented implicitly
but later explicitly in a tree. Subscriptions can be outdated after a branch is moved.
Failures in thorough change detection itself can occur.

Although ordinary state testing (checklist, one at a time: works for nominal, caution, danger)
would validate the implementation, these transition problems can hide latent bugs.

In run-throughs of pseudorandom sequences with pseudorandom fault injections, failures were
detected universally and immediately.  A short and simple script (comparing the JSON of which
the GUI was ``aware'' to the most-recent JSON) quickly identified bugs, with timestamps.

Whether run in advance, or applied to a logged history which has not yet been debugged, the
awareness-based, timestamped failures made it very easy to focus on finding root causes without
needing hours of manual, side-by-side inspection.  Figs.~\ref{fig-aware-good},\ref{fig-aware-bad} show
the reconstructed JSON during correct vs. faulty rendering.  Mismatched JSON immediately points
to GUI failures.

\subsection{Live Self-Validation using Awareness}

Although not specifically a required goal, the OD evaluation is compact and fast enough to
deploy with the application.  A version of our implementation included 1 Hz live-snapshots to
validate against incoming JSON.  An ``shame'' indicator was added to the GUI, to be shown when
validation failed.

Under fault injections (\S\ref{subsec-debug}), we observed that the self-aware GUI would show
shame within a second of fault injection.  (Demo also included in github.)

The app was instrumented to include fault injection (\S\ref{subsec-debug}). For example, an
error was simulated whereby state-change detection failed for transitions between {\em caution}
and {\em danger}. By being {\em aware} of the JSON using the interpreter-validator, live, the
GUI promptly showed the shame indicator within a second of fault injection. (This demonstration
is also included in github.)

Figs.~\ref{fig-correct-gui} shows satisfactory GUI display (correct backend GUI is derivable
from the snapshot), while fig.~\ref{fig-inject-shame} shows in-flight detection of mismatched
JSON, leading to a self-aware display of shame.

The CoreML object detector net was 61 MB in size, close to double the average 34 MB iOS app size
(circa 2020).  With a current 200 MB cap on deployed apps, self-awareness does not present an
impossible drawback for sampling deployments, although it increases battery drain.  Full online
awareness of a larger backend JSON tree would become prohibitive for store-deployed apps but
less so for dedicated devices such as vehicle dashboards.


\section{Cryptography} \label{s-crypto}

Imagine voters sending RSA-encrypted votes to a ballot counting center. Although Bob doesn't
know Alice's vote, Bob can copy Alice's vote by copying her ciphertext, in a naive setting.
Strictly speaking, this is a voting violation that doesn't occur with ideal paper ballots.
(Although this might not seem extreme, there are worse breakdowns.)

Historically, Shannon's foundational analysis of one-time pad encryption security focused on
privacy of the ciphertext \citep{sha49}. Other properties later became important - such as
integrity or authentication - but the focus remained on analyzing lists of ciphertext
properties, with surprisingly subtle pitfalls.

Decades later, a different paradigm emerged: a ``real'' vs ``ideal'' approach, focusing instead
on interactions between parties in two different settings \citep{bea91-resil,can01-uc}.  An
attacker in a ``real'' setting (using a deployed cryptosystem) must be mapped to an attack in
an ``ideal'' setting (e.g.  a trusted party or axiomatically-secure channel).  An implication
of this approach is that {\em successfully} encrypting a message requires reconstructibility of
the cleartext. Now Bob can't bluff.

Formally, this boils down to exhibiting a ``simulator'' who provides a mockup of the real world
to an adversary, while itself playing in the ``ideal'' protocol, where axiomatically-secure
channels are postulated.  As in a Turing test for intelligence, if it's infeasible to tell the
difference between real-world and ideal-world attacks, then whatever an adversary can achieve
using real life cryptosystems must also therefore be achievable in an ideal setting. In this
formalism of what it means to be secure, the simulator must \textit{extract an actual message}
to convey on the ideal channel.

Our approach and our definitions \ref{def-aware},\ref{def-aware-filter} of awareness are
motivated by this paradigm.  Where a crypto simulator demonstrates an adversary's awareness of
the ciphertext (by building the cleartext sent in an ideal channel), our mapping from rendering
to backend JSON is a demonstration that the renderer is {\em aware} of the backend JSON.

The fidelity of the synthesized JSON, like the fidelity of a synthesized cleartext, is core to
this approach. The formal and practical benefits make reverse-engineering the JSON a challenge
worth pursuing.

While crypto validation admits a wide range of adversaries (poly-time attackers), our adversary
class is limited to collections of challenging test sequences.  The generation of
broad-coverage test sequences remains important (and an independent task), but fortuitously,
unlike arbitrary crypto adversaries, we can increase the validation level by scaling up from
simpler to broader scenarios.


\section{Closing Remarks} \label{s-summary}

We proposed a GUI validation approach focusing on constructive awareness of the backend
communications rather than on various properties of the rendering itself and how they change
over time.  We showed an efficient and concrete single-platform implementation, demonstrating
that disciplined TDD for GUI development can be accomplished without ``cheating''
(\textit{i.e.} without needing an initial implementation to generate golden test images). The
methodology is cross-platform and independent of the rendering technology -- it does not need
to capitalize on platform-specific aspects like accessibility features or DOM/UIKit structures.

Generating comprehensive test sequencing remains a critical part of validation but is
independent of this direction. In fact, our approach supports much greater flexibility of
testing based on live, empirical, unseen sequences -- it is not confined to comparisons to
prepackaged images from fixed, pre-established transition-coverage collections.

Post-testing can be supported by training with automatically-labeled images from a production
implementation or MVP -- using labels from the backend JSON.  Testing is bootstrapped from
verified renderings of well-covered \textit{states}, decoupled from efforts to cover all
\textit{state transitions}. This style of migration is arguably more robust and objective than
golden-image replacement.

Several directions lead from here. Many IDEs provide templates for test generation; here,
practical toolkits to automate and streamline the full design-to-test route can help, including
the post-testing auto-training approach.  As in crypto, the composition of awareness of
separately-vetted subtrees needs formal support, as there are some edge-case pitfalls. Some
otherwise-idle challenges gain a practical application -- for example, reverse engineering GPS
from diverse map renderings.

\clearpage

\nocite{*}
\bibliographystyle{plain}  
\bibliography{2020-appl-aware-v1}

\end{document}